\documentstyle[prl,aps,epsfig]{revtex}

\begin{document}

%%%%%%%%%%%%%%%%%%%%%%%%%%%%%%%%%%%%%%%%%%%%%%%%%%%%%%%%%%%%%%%%%%%%%
%%%%%%%%%%%   TITLE PAGE            %%%%%%%%%%%%%%%%%%%%%%%%%%%%%%%%%
%%%%%%%%%%%%%%%%%%%%%%%%%%%%%%%%%%%%%%%%%%%%%%%%%%%%%%%%%%%%%%%%%%%%%

\title{ The Markovian metamorphosis of a simple turbulent
        cascade model }

\author{
Jochen Cleve$^{1,2}$ and Martin Greiner$^{1,2}$
}

\address{$^1$Institut f\"ur Theoretische Physik, Technische Universit\"at,
             D--01062 Dresden, Germany }
\address{$^2$Max-Planck-Institut f\"ur Physik komplexer Systeme, 
             N\"othnitzer Str.\ 38, D--01187 Dresden, Germany }

\date{20.03.2000}

\maketitle

\begin{abstract}
Markovian properties of a discrete random multiplicative cascade model 
of log-normal type are discussed. After taking small-scale resummation
and breaking of the ultrametric hierarchy into account, 
qualitative agreement with Kramers-Moyal coefficients, recently deduced
from a fully developed turbulent flow, is achieved.
\end{abstract}

\vspace*{2cm}


\vspace*{4cm}

\noindent
CORRESPONDING AUTHOR: \\
Martin Greiner \\
Max-Planck-Institut f\"ur Physik komplexer Systeme \\
N\"othnitzer Str.\ 38 \\ 
D--01187 Dresden, Germany  \\
tel.: 49-351-871-1218 \\
fax:  49-351-871-1199 \\
email: greiner @ mpipks-dresden.mpg.de

\newpage
%----------------------------------------------------------------------
%----------------------------------------------------------------------

Phenomenological modelling of the energy cascade in fully developed 
turbulence has a long tradition \cite{FRI95,BOH98}. As representatives of 
the multifractal approach random multiplicative branching processes
\cite{NOV71,MAN74,MEN91} mimic the redistribution of energy flux from
the large, integral length scale $L$ down to the small, dissipative
scale $\eta$ and focus on the scaling aspect of the surrogate energy
dissipation field, extracted from measured velocity time series. In a
particular simple model version, which is only one-dimensional and binary
discrete, a domain of length $r_j=L/2^j$ is split into two subdomains
of equal length $r_{j+1}$ and the energy flux density $\varepsilon(r_j)$
of the parent domain is non-uniformly redistributed by assigning a 
left/right multiplicative weight $q_{{\cal L}/{\cal R}}$ 
to the left/right subdomain:
$\varepsilon_{\cal L}(r_{j+1}) = q_{\cal L} \varepsilon(r_j)$,
$\varepsilon_{\cal R}(r_{j+1}) = q_{\cal R} \varepsilon(r_j)$.
The multiplicative weights are drawn from a 
scale-independent symmetric probabilistic 
splitting function $p(q_{\cal L},q_{\cal R}) = p(q_{\cal R},q_{\cal L})$, 
are $\langle{q_{\cal L}}\rangle=\langle{q_{\cal R}}\rangle=1$ 
on average and are completely uncorrelated to multiplicative weights 
from all other branchings differing in scale and position. At an
intermediate length scale $\eta{\leq}r_j{\leq}L$, corresponding to $j$
cascade steps, the local bare field density
$\varepsilon(r_j) =  q_1 q_2 \cdots q_j$
is a product of $j$ independent multiplicative weights, where 
$\varepsilon(r_0) = 1$ has been chosen for simplicity. Upon taking the 
logarithm, the product turns into a summation over independent and 
identically distributed random variables:
%----------------------------------------------------------------------
\begin{equation}
\label{one}
  \ln\varepsilon(r_j)
    =  \ln{q_1} + \ln{q_2} + \ldots +  \ln{q_j}
       \; .
\end{equation}
%----------------------------------------------------------------------
Introducing 
$y_j = y(l_j) = \ln\varepsilon(r_j) - \langle{\ln\varepsilon(r_j)}\rangle$
as new field variable and $l_j = \ln{(L/r_j)} = j\ln{2}$ as a logarithmic
scale, it is straightforward to derive
%----------------------------------------------------------------------
\begin{equation}
\label{two}
  \frac{\Delta{y}}{\Delta{l}} 
    =  \frac{y_{j+1} - y_j}{l_{j+1} - l_j}
    =  \frac{1}{\ln{2}} 
       \left( \ln{q_{j+1}} - \langle{\ln{q}}\rangle \right)
    =  \sqrt{ 
       \frac{\langle{\ln^2{q}}\rangle-\langle{\ln{q}}\rangle^2}{2\ln{2}}
       } \xi_{j+1}
\end{equation}
%----------------------------------------------------------------------
from (\ref{one}) for parent/daughter field variables. The last step,
leading to the stationary Gaussian-white noise ``random force'' $\xi_j$
with normalisation 
$\langle \xi_j \xi_{j^\prime} \rangle = \frac{2}{\ln{2}}\delta_{jj^\prime}$
\cite{RIS89}, only holds for a splitting function 
$p(q_{\cal L},q_{\cal R})=p(q_{\cal L})p(q_{\cal R})$ 
of log-normal type, where
%----------------------------------------------------------------------
\begin{equation}
\label{three}
  p(q)
    =  \frac{1}{\sqrt{2\pi}\sigma{q}}
       \exp\left( 
       -\frac{1}{2\sigma^2} \left( \ln{q}+\frac{\sigma^2}{2} \right)^2
       \right)
       \; .
\end{equation}
%----------------------------------------------------------------------
The Langevin equation (\ref{two}) represents a discrete
Markov process, evolving from large to small scales with zero drift term
$D^{(1)}=0$ and constant diffusion term 
$D^{(2)}=(\langle{\ln^2{q}}\rangle-\langle{\ln{q}}\rangle^2)/2\ln{2}$.

In several ways this thinking seemingly contradicts the results
deduced from a large-Reynolds number helium jet experiment 
\cite{NAE97,MAR98}: 
from the coarse-grained one-dimensional surrogate energy dissipation
field
%----------------------------------------------------------------------
\begin{equation}
\label{four}
  \overline{\varepsilon}(x,r)
    =  \frac{1}{r} \int_{x-{r \over 2}}^{x+{r \over 2}}
       \varepsilon(x^\prime,\eta) {\rm d}x^\prime
       \; ,
\end{equation}
%----------------------------------------------------------------------
entering into the transformed variable
%----------------------------------------------------------------------
\begin{equation}
\label{five}
  \overline{y}(x,l) 
    =  \ln\overline{\varepsilon}(x,r)
       - \langle \ln\overline{\varepsilon}(x,r) \rangle
       \; ,
\end{equation}
%----------------------------------------------------------------------
the Kramers-Moyal coefficients
%----------------------------------------------------------------------
\begin{equation}
\label{six}
  D^{(n)}(\overline{y}(l))
    =  \lim_{\Delta{l}\rightarrow{0}} \frac{1}{n!\Delta{l}}
       \int \left[ \overline{y}(l+\Delta{l})-\overline{y}(l) \right]^n
       p(\overline{y}(l+\Delta{l})|\overline{y}(l))
       {\rm d}\overline{y}(l+\Delta{l})
\end{equation}
%----------------------------------------------------------------------
have been determined to yield
%----------------------------------------------------------------------
\begin{eqnarray}
\label{seven}
  D^{(1)}(\overline{y})
    &=&  \gamma\overline{y}
         \qquad\qquad (\gamma\approx{0.21})
         \nonumber \\
  D^{(2)}(\overline{y})
    &=&  D
         \qquad\qquad (D\approx{0.03})
         \\
  D^{(n\geq{3})}(\overline{y})
    &\approx&  0
         \; .
         \nonumber
\end{eqnarray}
%----------------------------------------------------------------------
This outcome suggests that the energy cascade in fully developed 
turbulence can be described by a scale-continuous Markovian
Ornstein-Uhlenbeck process, which differs from the discrete random 
multiplicative branching picture in two ways:
scale-continuous evolution with a linear drift and constant diffusion term
as opposed to a scale-discrete evolution with a zero drift and constant
diffusion term. 
--
However, the comparison between these two, apparently contradicting 
pictures is not well taken and is not as straightforward as anticipated. 
So far it has been like comparing a caterpillar with a butterfly. Now, 
we will initiate the metamorphosis in two steps.

Step one has to do with the distinction between a bare and a dressed 
field \cite{SCH87}. Since fully developed turbulence is a three-dimensional
process, the redistribution of energy flux from larger to smaller scales
should be conserved in three dimensions, as long as the dissipative scale
$\eta$ is not reached. Of course this does not hold once the process is
looked at in only one dimension, which is done in one-point time-series
measurements of one component of the velocity field and from which the
one-dimensional surrogate energy dissipation field is extracted. For the
simple binary discrete random multiplicative branching process this 
implies that the splitting function 
$p(q_{\cal L},q_{\cal R}) \approx p(q_{\cal L}) p(q_{\cal R})$ more or less
factorises, where $p(q)$ should be a positively skewed distribution
limited to the support $0{\leq}q{\leq}q_{\rm max}$ with 
$q_{\rm max}\approx{2^3}$ \cite{JOU00}. In this respect, the log-normal
distribution (\ref{three}) with the realistic parameter $\sigma=0.42$,
reproducing observed lowest-order scaling exponents
$\langle \varepsilon^n(r_j) \rangle{\sim}(L/r_j)^{\tau(n)}$
and observed multiplier distributions, represents a fair candidate. With
a factorised splitting function, where 
$\langle q_{\cal L}{+}q_{\cal R} \rangle =2$ 
only holds on average, we have to distinguish between the bare field 
(\ref{one}), which is evolved from the large scale $L$ down to the
intermediate scale $r_j$, and the dressed field
%----------------------------------------------------------------------
\begin{equation}
\label{eight}
  \overline{\varepsilon}(r_j)
    =  \varepsilon(r_j) (1+\Delta(r_j))
       \; ,
\end{equation}
%----------------------------------------------------------------------
which has been evolved from $L$ all the way down to the dissipative
scale $\eta$ and then again resummed up to the intermediate scale $r_j$.
The two fields differ by a small-scale resummation factor 
$(1+\Delta(r_j))$, which is equal to one only on average. As the 
experimental analysis (\ref{four}) corresponds to the dressed field, we
also need to employ the dressed field for the random multiplicative branching
process, in order to make a fairer comparison between model and data 
results.

For a truly fair comparison between model and data results we have to
call for an additional, second step: since the binary discrete random
multiplicative branching process is organised hierarchically in
one-dimensional space, the underlying ultrametric does not allow for
spatially homogeneous observables right away. This has been noted only
recently \cite{JOU00,GRE97,JOU99a} and a simple scheme has been suggested,
breaking the ultrametric hierarchy and restoring spatial homogeneity. It 
builds a long chain of independent cascade field realisations, each of
length $L$, randomly places the observational interval of length
$\eta{\leq}r{\leq}L$ within this chain and samples over these random 
placings.

These two steps have been decisive for the correct interpretation
\cite{JOU00,JOU99a} of the observed multiplier phenomenology
\cite{SRE95,PED96}: the small-scale resummation of step one explains
the scale-independent multiplier distributions as fixed-point 
distributions and step two is in charge for producing the correct
correlations between multipliers.
--
Since the Kramers-Moyal coefficients (\ref{six}) can be understood as
moments of logarithmic multipliers, defined for an infinitesimal
scale step, we might already begin to speculate here that steps one
and two might also be in charge for turning the caterpillar (\ref{two})
into the butterfly (\ref{seven}).

We will now test this speculation by numerical simulation of the 
binary discrete random multiplicative branching process with the 
factorised splitting function (\ref{three}) of log-normal type 
($\sigma = 0.42$). A chain of $N_L=10^6$ independent cascade
realisations is constructed, where each realisation has been obtained
after $J=10$ binary cascade steps; consequently the length of the total
chain amounts to $10^6 L = 1.28{\cdot}10^9 \eta$.

At first we test the Markov property in general. The conditional
probability distributions
$p(\overline{y}(l_2)|\overline{y}(l_1))$
with centred intervals $l_2>l_1$ is sampled over $e^{l_1} N_L$ random
placings within the long chain of cascade realisations. It is found 
that these conditional probability distributions fulfill the
Chapman-Kolmogorov equation
$p(\overline{y}(l_3)|\overline{y}(l_1))
 = \int p(\overline{y}(l_3)|\overline{y}(l_2))
        p(\overline{y}(l_2)|\overline{y}(l_1))
        {\rm d}\overline{y}(l_2)$
close to perfectly in the scale range $\eta{\ll}l{\leq}L$. This is a
necessary and almost sufficient validation for this branching process
to appear Markovian \cite{RIS89}.

Without any loss of generality the Kramers-Moyal coefficients 
(\ref{six}) are only calculated at binary scales $r_j$, but again
sampled over randomly chosen $x$-values within the long cascade chain.
Convergence is tested by letting the positive integer number 
$m{\rightarrow}1$ in the centred daughter interval of length
$r_j{-}\Delta{r_j} = r_j{-}2m\eta$. Good convergence is achieved for
$0{\leq}j{\leq}5$, i.e.\ the upper part of the cascade inertial range,
whereas for $6{\leq}j{\leq}J=10$ convergence has been found to be
unsatisfactory since $\eta{\ll}r_j$ is no longer fulfilled. Consequently, 
in the following we will only show results for the former scale range. 
--
The first Kramers-Moyal coefficient $D^{(1)}(\overline{y},l)$ is
illustrated in Fig.\ 1a at binary scales $j=1,3,5$. It is not constant
zero anymore; now it is linear in $\overline{y}(l_j)$. Fitting the
parametrisation
$D^{(1)}(\overline{y},l) = \gamma_0(l) + \gamma(l)\overline{y}$
yields $\gamma_0(l)=0$ within simulation error bars and a positive,
slightly scale-dependent drift coefficient $\gamma(l)$ with values
listed in Tab.\ 1a. The latter decreases from a value $0.19$ at 
$L=2^{9}\eta$ to a value $0.08$ at $r_5=2^5\eta$ and agrees with the
experimentally deduced value (\ref{seven}) within a factor of 1.1-2.5. 
The second Kramers-Moyal coefficient $D^{(2)}(\overline{y},l)$, depicted
in Fig.\ 1b, turns out to be almost constant and almost scale-independent;
fitted values of the parametrisation
$D^{(2)}(\overline{y},l) = D(l) + d_1(l)\overline{y}$ can be found in 
Tab.\ 1a. Compared with $D=0.127$ of the caterpillar thinking (\ref{two})
it is reduced by about a factor of 3, but is still about a factor of
$1.4$ above the experimentally given result (\ref{seven}). Also in 
qualitative agreement with the experimental results, higher-order
Kramers-Moyal coefficients are close to zero: $D^{(3)}(\overline{y},l)$
and $D^{(4)}(\overline{y},l)$ are of the order $10^{-3}$ and $10^{-4}$,
respectively. Here we might evoke Pawulas theorem \cite{RIS89} to 
conclude that, after taking steps one and two into account,
the binary discrete random multiplicative branching process
of log-normal type appears as a scale-continuous Markovian 
Ornstein-Uhlenbeck process with Kramers-Moyal coefficients
%----------------------------------------------------------------------
\begin{eqnarray}
\label{nine}
  D^{(1)}(\overline{y})
    &=&  \gamma(l)\overline{y}
         \qquad\qquad (0.2\geq\gamma(l)\geq{0.1})
         \nonumber \\
  D^{(2)}(\overline{y})
    &=&  D
         \qquad\qquad\quad\; (D\approx{0.04})
         \\
  D^{(n\geq{3})}(\overline{y})
    &\approx&  0
         \; .
         \nonumber
\end{eqnarray}
%----------------------------------------------------------------------
This result is in nice qualitative agreement with the experimental
observation (\ref{seven}).

The result (\ref{nine}) has been obtained by taking both steps, one and
two, of the metamorphosis
into account. Step one with its small-scale resummation is
definitely responsible for the transition from the scale-discrete
evolution of the bare field to the scale-continuous Markov description 
for the dressed field. What are then the implications of step two, i.e.\
the breaking of the ultrametric cascade hierarchy to restore spatial
homogeneity? In order to clarify this point, we restrict the sampling
of the Kramers-Moyal coefficients only to the hierarchical positions
$x_m=(m+0.5)r_j$ with integer $0{\leq}m{<}2^j{\cdot}N_L$
within the long chain of cascade configurations.
For these positions, the integration interval
of length $r_j$, entering into (\ref{four}), perfectly matches an
intermediate interval of the bare cascade evolution. Results for the 
first and second Kramers-Moyal coefficients are listed in Tab.\ 1b.
The first coefficient, which we now denote with a tilde, is again
found to be of the form
$\tilde{D}^{(1)}(\overline{y},l) = \tilde{\gamma}(l)\overline{y}$.
Note however, that the drift coefficient $\tilde{\gamma}$ is negative
and that its modulus is about a factor of 3-7 less when compared with 
$\gamma$. This demonstrates that small-scale resummation alone
already introduces a weak linear drift term, but breaking of the 
ultrametric hierarchy is very necessary to change its sign and to bring
it to the correct order of magnitude. Also the second Kramers-Moyal 
coefficient 
$\tilde{D}^{(2)}(\overline{y},l) 
 = \tilde{D}(l) + \tilde{d}_1(l)\overline{y}$
is affected by leaving out step two, but only weakly:
again it shows a small scale-dependence, is almost constant for a given
scale and, once compared with $D(l)$, is reduced by about a factor of
1.2-1.7. Note also, that it is mainly small-scale resummation, which drives
the diffusion coefficient away from the caterpillar thinking (\ref{two})
with $D=0.127$.

We conclude: 
small-scale resummation and breaking of the ultrametric hierarchy
initiate the Markovian metamorphosis of a discrete random multiplicative
branching process as they turn the caterpillar, a discrete, Gaussian
white-noise evolution in scale, into a butterfly, that is an effective
scale-continuous Ornstein-Uhlenbeck description, the latter being in
qualitative agreement with the experimental observations. At least on
a qualitative level, there is no conflict between the experimentally
observed energy cascade in fully developed turbulence and random
multiplicative branching processes; this statement is further supported
by recent work on multiplier distributions \cite{JOU00,JOU99a,JOU99b}.
--
Several points have to be considered in order to possibly achieve an
even better quantitative agreement between such model results and 
experimental observations: 
sensitivity on the generator of random multiplicative branching processes,
i.e.\ on scale-discrete or scale-continuous implementations and on the
choice of splitting function, as well as on finite-size, large-scale and
dissipative effects; work in these directions is in progress
\cite{CLE00}. Also data results for larger Reynolds numbers would
be highly appreciated.

\newpage
%%%%%%%%%%%%%%%%%%%%%%%%%%%%%%%%%%%%%%%%%%%%%%%%%%%%%%%%%%%%%%%%%%%%
%%%%%%%%%%%%%%%%%%%%%%%%%%%%%%%%%%%%%%%%%%%%%%%%%%%%%%%%%%%%%%%%%%%%

\vspace*{8cm}

%%%%%%%%%%%%%
% Table 1   %
%%%%%%%%%%%%%
\begin{table}[b]
\centering
\begin{tabular}{|c|ccc|ccc|}
%        & \multicolumn{3}{c|}{(a)}
%        & \multicolumn{3}{c|}{(b)} \\
      & \multicolumn{1}{l}{(a)} & & & \multicolumn{1}{l}{(b)} & & \\
  $j$ & $\gamma$ & $D$ & $d_1$ & 
          $\tilde{\gamma}$ & $\tilde{D}$ & $\tilde{d_1}$\\[0.5ex] \hline
    0 & 0.154 & 0.042 & -0.004 & -0.178 & 0.036 & 0.004 \\
    1 & 0.193 & 0.042 & -0.004 & -0.077 & 0.034 & 0.002 \\
    2 & 0.163 & 0.040 & -0.003 & -0.052 & 0.031 & 0.001 \\
    3 & 0.133 & 0.040 & -0.002 & -0.034 & 0.029 & 0.001 \\
    4 & 0.105 & 0.040 & -0.002 & -0.021 & 0.027 & 0.001 \\
    5 & 0.084 & 0.043 & -0.002 & -0.012 & 0.025 & 0.001
\end{tabular}
\caption{
First and second Kramers-Moyal coefficients,
$D^{(1)}(\overline{y},l_j) = \gamma(l_j) \overline{y}$ and 
$D^{(2)}(\overline{y},l_j) = D(l_j) + d_1(l_j)\overline{y}$,
after small-scale resummation and (a) with / (b) without breaking
of the ultrametric hierarchy.
}
\end{table}

\newpage
%%%%%%%%%%%%%
% FIGURE 1  %
%%%%%%%%%%%%%
\begin{figure}
\begin{centering}
\epsfig{file=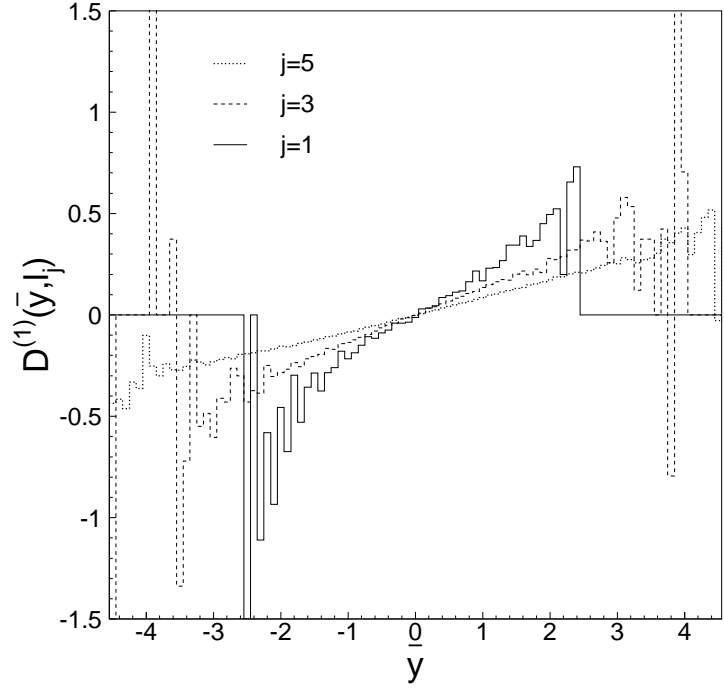,width=10cm}
\epsfig{file=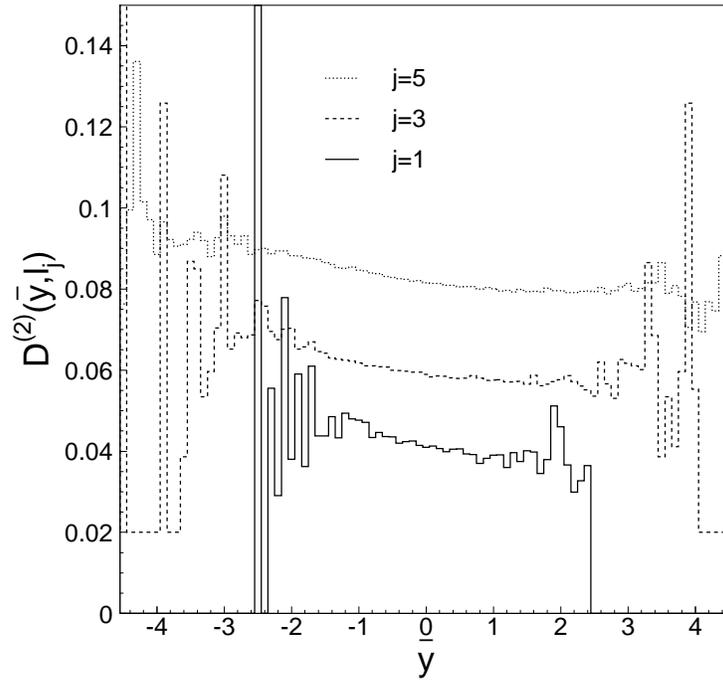,width=10cm}
\caption{
First (a) and second (b) Kramers-Moyal coefficients, 
obtained after small-scale resummation and breaking of the ultrametric
hierarchy. The $D^{(2)}$-values for $j=3$ and $5$ have been artificially
shifted by $0.02$ and $0.04$, respectively.
} 
\end{centering}
\end{figure}

\end{document}